\documentclass[10pt,conference]{IEEEtran}
\IEEEoverridecommandlockouts

\ifCLASSINFOpdf
\else
\fi

\usepackage{cite}
\usepackage{amsmath,amssymb,amsfonts}

\usepackage{graphicx}
\usepackage{textcomp}
\usepackage{xcolor}
\usepackage{multirow}
\usepackage{bm}
\usepackage{booktabs}
\usepackage{tabularx}
\usepackage[caption=false]{subfig}
\usepackage{enumitem} 
\usepackage{tikz}
\usetikzlibrary{shapes.geometric, arrows}
\usepackage[linesnumbered, ruled, noend]{algorithm2e}

\allowdisplaybreaks[4]

\begin{document}

\title{UAV-Assisted Joint Data Collection and Wireless Power Transfer for Batteryless Sensor Networks}

\author{\IEEEauthorblockN{
        Wen Zhang\IEEEauthorrefmark{2},
        Aimin Wang\IEEEauthorrefmark{2},
        Geng Sun\IEEEauthorrefmark{2}\IEEEauthorrefmark{3}\IEEEauthorrefmark{1},
        Jiahui Li\IEEEauthorrefmark{2}\IEEEauthorrefmark{1},
        Jiacheng Wang\IEEEauthorrefmark{3},
        Changyuan Zhao\IEEEauthorrefmark{3},
        Dusit Niyato\IEEEauthorrefmark{3}
    }
    
    \IEEEauthorblockA{\IEEEauthorrefmark{2}College of Computer Science and Technology, Jilin University, Changchun 130012, China}
    \IEEEauthorblockA{\IEEEauthorrefmark{3}College of Computing and Data Science, Nanyang Technological University, Singapore 639798, Singapore}
    \IEEEauthorrefmark{1}Corresponding authors: Geng Sun and Jiahui Li}

\markboth{Journal of \LaTeX\ Class Files,~Vol.~14, No.~8, August~2021}%
{Shell \MakeLowercase{\textit{et al.}}: A Sample Article Using IEEEtran.cls for IEEE Journals}

\maketitle

\IEEEtitleabstractindextext{%
\begin{abstract}
The development of wireless power transfer (WPT) and Internet of Things (IoT) offers significant potential but faces challenges such as limited energy supply, dynamic environmental changes, and unstable transmission links. This paper presents an unmanned aerial vehicle (UAV)-assisted data collection and WPT scheme to support batteryless sensor (BLS) networks in remote areas. In this system, BLSs harvest energy from the UAV and utilize the harvested energy to transmit the collected data back to the UAV. The goal is to maximize the collected data volume and fairness index while minimizing the UAV energy consumption. To achieve these objectives, an optimization problem is formulated to jointly optimize the transmit power and UAV trajectory. Due to the non-convexity and dynamic nature of the problem, a deep reinforcement learning (DRL)-based algorithm is proposed to solve the problem. Specifically, this algorithm integrates prioritized experience replay and the performer module to enhance system stability and accelerate convergence. Simulation results demonstrate that the proposed approach consistently outperforms benchmark schemes in terms of collected data volume, fairness, and UAV energy consumption.
\end{abstract}

}

\maketitle
%

\IEEEdisplaynontitleabstractindextext
\IEEEpeerreviewmaketitle


%
%
\section{Introduction} 
\label{sec:introduction}
\par
Internet of Things (IoT) proliferation has revolutionized data collection, with billions of devices generating information for decision-making. However, this massive deployment faces significant energy constraints, which are worsened by power-intensive applications in challenging environments. Wireless power transfer (WPT) offers a promising solution to replenish device energy\cite{Li2025}. However, conventional WPT deployments struggle in extreme environments like high-temperature or hazardous industrial areas. Fixed infrastructure cannot be safely installed there, and conventional batteries degrade rapidly.
\par
These limitations have spurred the development of batteryless sensors (BLS). BLS harvest energy from radio frequency signals \cite{Cai2023}, which provide sustainability, reduced maintenance needs, and capability in environments where battery access is impractical. Nevertheless, BLS deployments in harsh settings encounter challenges in power delivery and data retrieval.
\par
Uncrewed aerial vehicles (UAVs) present a solution. Specifically, UAVs function as mobile platforms interacting with ground-based sensors. Furthermore, they overcome fixed infrastructure limitations. Additionally, UAVs integrate WPT modules as aerial transmitters, which provide power to BLS through directional radio frequency (RF) transmission while the UAVs collect sensor data.
\par
However, current research focuses on self-powered sensors and overlooks constraints of BLS networks. Moreover, existing frameworks employ single-objective approaches that optimize either data collection or energy consumption, whereas BLS networks necessitate multi-objective optimization balancing collected data volume, fairness index, and UAV energy consumption. Therefore, effective UAV-assisted BLS systems require three key components, UAV transmit power control for data collection under time-varying channels, trajectory optimization balancing fairness and UAV energy consumption, and online decision making to handle uncertainty. Accordingly, the main contributions include:

\begin{itemize}
    \item \textit{UAV-Assisted Joint Data Collection and WPT System for BLS Networks}: We investigate a UAV-assisted joint data collection and WPT system for BLS networks. Specifically, UAVs utilize WPT technology to energize batteryless sensor nodes while  collecting data from them. Such scenarios are common and practical in remote areas, such as desert environments.
    \item \textit{Joint Optimization of UAV Transmit Power and Trajectory}: We formulate a joint optimization problem to coordinate the transmit power and trajectory of the UAV, which aims to maximize the collected data volume and fairness index while minimizing the UAV energy consumption. This problem is challenging due to the high real-time requirements and the dynamic nature of the environment.
    \item \textit{Enhanced DRL Approach}: We propose a SAC-PP, which is a DRL-based algorithm to solve the formulated optimization problem. Specifically, SAC-PP integrates prioritized experience replay (PER) and the performer module to enhance the system stability and accelerate the convergence.
    \item \textit{Performance Evaluation}: Simulation results demonstrate that the proposed SAC-PP algorithm outperformed other baseline methods in terms of collected data volume, fairness, and  energy consumption.
\end{itemize}
\par The structure of this paper is outlined as follows. Section II describes the models and problem setup. Section III details the proposed SAC-PP algorithm. Section IV presents the simulation results, and Section V provides concluding remarks.

%
\begin{figure}[!t]
  \centering
  \includegraphics[width=3.5in]{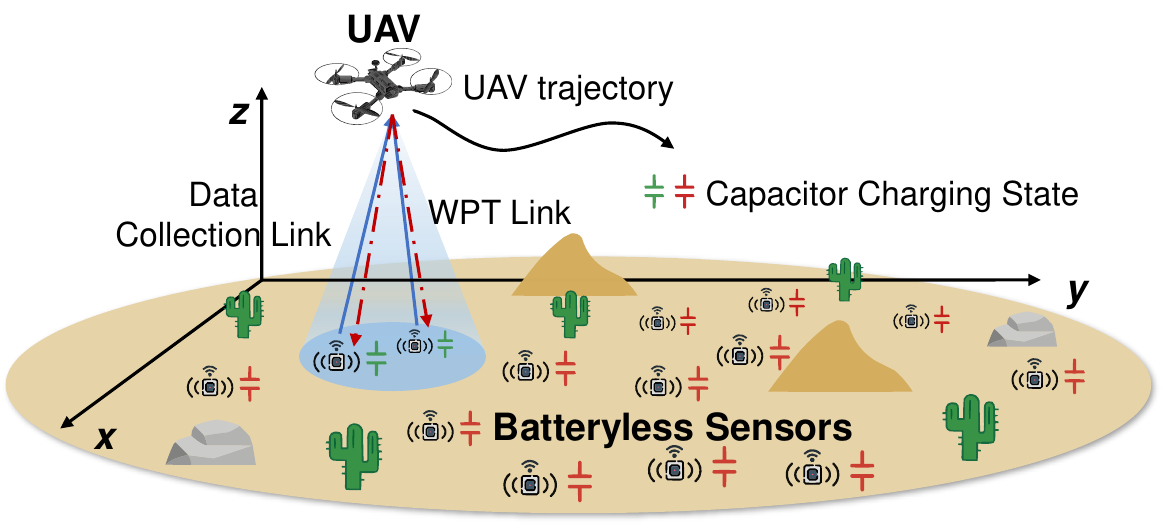}
  \caption{UAV-assisted joint data collection and WPT system for a BLS network. The UAV first charges BLSs within its distance threshold through WPT. BLSs perform sensing and monitoring tasks, and utilize the harvested energy to transmit the collected data back to the UAV using a multiple access scheme.}
  \label{fig:network-model}
\end{figure}

%
%
\section{System Model} 
\label{sec:models_and_preliminaries}

\par In this section, we outline the models and preliminaries of this work.

%
\subsection{Network Model} 

\par As shown in Fig.~\ref{fig:network-model}, a UAV-assisted joint data collection and WPT system operates in remote areas such as a desert environment. The monitoring area $A_m$ contains many BLSs. Despite advantages of wireless sensor networks, battery use limits network lifespan and causes environmental pollution \cite{Cai2023}. Batteryless sensors harvest energy from external sources like RF signals and store energy in supercapacitors for short-term, eco-friendly use, which overcomes battery challenges in remote areas \cite{Cai2023}. Due to low maintenance and reliability in harsh environments without battery replacement, a group of BLSs equipped with supercapacitors collects data for environmental monitoring. BLSs form the set $\mathcal{W} = \{ i \mid i = 1, 2, \ldots, N_{SN} \}$. Challenges of installing fixed power infrastructure and base stations in extreme environments like high-temperature or hazardous industrial areas lead to deployment of a quadrotor UAV, which delivers power to BLSs via WPT and collects transmitted data. Each sensor and UAV uses an omnidirectional antenna, while positions of BLSs remain predetermined and static. The range of the UAV for data collection and WPT stays limited by the distance threshold $D_{max}$.

\par In each mission cycle, the BLSs need to be powered and upload their sensed data for data backup and security. For clarity, time is partitioned into T slots, each with a duration of $t_d$, and denoted by $\mathcal{T}=\{t \mid t=1,2, \ldots, T\}$. Due to limited energy capacity, BLSs can only support low-power sensing but not energy-intensive transmission, so they lack power for transmission until after WPT \cite{Cai2023}. Moreover, each BLS has a certain probability of generating new data collection requests in each time slot. The heterogeneous spatial distribution of BLSs, coupled with the dynamic nature of data requests and the variable energy consumption patterns of the UAV, poses significant challenges to real-time trajectory optimization. To address these challenges, the UAV dynamically adjusts its flight path to simultaneously engage with multiple BLSs within a single time slot, thereby coordinating WPT and data transmission. Specifically, the UAV delivers downlink WPT to replenish BLS energy reserves, while employing an orthogonal frequency division multiple access (OFDMA) scheme for bandwidth allocation during uplink data transmission. Note that our method does not depend on a specific protocol and can be easily extended to alternative multiple access schemes, such as TDMA or LoRa, by introducing appropriate routing and networking protocols.

\par Considering a three-dimensional (3D) Cartesian coordinate system, the position of the $i$-th sensor in the monitoring area $A_{m}$ is represented as $\mathbf{c}_i^{SN} = (x_i^{SN}, y_i^{SN}, 0)$. Since the duration of each time slot $t_d$ is sufficiently small, the position of UAV is regarded as fixed within any time slot $t$ and is expressed as $\mathbf{c}^{U}[t] \triangleq (x^U[t], y^U[t], z^U), \forall t \in \mathcal{T}$. To maintain balanced data acquisition in the BLS network and optimize UAV energy consumption, we subsequently formulate key operational models, including the WPT model, wireless data collection model, and UAV mobile and energy cost models.

%
\subsection{WPT Model} 
\par During time slot $t$, the UAV provides wireless power to BLSs within its coverage area by using a transmit power of $P_\mathrm{w}^U[t]$. Let $d_i^{SNU}[t]$ denote the distance between the UAV and the $i$-th sensor node at time slot $t$. If this distance is within the threshold $D_{\text{max}}$, sensor $i$ remains operational. The received power at sensor $i$ is given by the Friis transmission equation~\cite{Panahi2024}, which is as follows:
\begin{equation}
P_i^r[t] = P_\mathrm{w}^{U}[t] \frac{G_T G_R \lambda^2}{(4 \pi)^2 (d_i^{SNU}[t])^\alpha},
\end{equation}
where $\lambda$ is the wavelength of the RF signal, $\alpha$ is the path loss exponent of the air-to-ground channel, and $G_T$ and $G_R$ are the antenna gains of the UAV and the sensor.

\par The rectifying antenna requires a minimum power $P_{\mathrm{min}}$ to activate, and $P_{\mathrm{max}}$ represents the saturation limit of the received power of the sensor. Specifically, at time slot $t$, the harvested power for sensor $i$ is given by
\begin{equation}
P_i^H[t](P_i^r[t]) =
\begin{cases}
    0, & P_i^r[t] < P_{\mathrm{min}}, \\
    \eta(P_i^r[t]) \cdot P_i^r[t], & P_{\mathrm{min}} \leq P_i^r[t] < P_{\mathrm{max}}, \\
    P_i^H[t](P_{\mathrm{max}}), & P_i^r[t] \geq P_{\mathrm{max}}.
\end{cases}
\end{equation}
where $P_i^H[t]$ converts the received power into usable direct current (DC) power. Using curve-fitting tools on empirical data, the function $\eta(\cdot)$ can be expressed as a polynomial for simplicity and accuracy. This work considers ideal antenna operation within the range $[P_{\mathrm{min}}, P_{\mathrm{max}}]$.

\par After completing the WPT, we proceed to discuss the communication model for data collection from the BLSs by the UAV.

%
\subsection{Wireless Data Collection Model} 
\par The UAV collects data from BLSs through a typical ground-to-air communication link. The channel coefficient between the UAV and sensor $i$ during time slot $t$ is denoted as $h_i[t] = \sqrt{\beta_{i}[t]} \tilde{h}_i[t]$, where $\beta_i[t]$ accounts for large-scale fading. The term $\tilde{h}_i[t]$ represents small-scale fading and is modeled as a complex random variable with $\mathbb{E}[|\tilde{h}_i[t]|^2] = 1$.

\par Generally, we consider an elevation angle-dependent probabilistic line-of-sight (LoS) model. This is due to the fact that the UAV can try to achieve a LoS link even in the presence of obstacles by moving. In this model, large-scale fading is viewed as a probabilistic variable influenced by the likelihood of LoS and NLoS conditions. Consequently, the large-scale channel coefficient $\beta_{i}[t]$ is given by
\begin{equation}
\beta_i[t] = 
\begin{cases}
    \beta_0 d_i^{SNU}[t]^{-\alpha}, & \text{LoS}, \\
    \kappa \beta_0 d_i^{SNU}[t]^{-\alpha}, & \text{NLoS},
\end{cases}
\end{equation}
where $\beta_0$ is the path loss at a reference distance, $\alpha$ is the path loss exponent, and $\kappa$ accounts for additional fading in NLoS environments. The probability of an LoS link is given by
\begin{equation}
P_{i, \text{LoS}}[t] = 1 / \left(1 + C \exp\left(-D\left(\theta_i[t] - C\right)\right)\right),
\end{equation}
where $C$ and $D$ are environment-dependent parameters, and $\theta_i[t] = (180/\pi) \sin^{-1}(z^U / d_i^{SNU}[t])$ is the elevation angle in degrees. The NLoS probability is $P_{i, \text{NLoS}}[t] = 1 - P_{i, \text{LoS}}[t]$. Thus, the expected channel power gain by averaging over both randomness can be expressed as follows:
\begin{equation}
\mathbb{E}[|h_i[t]|^2] = P_{i, \text{LoS}}[t] \beta_0 d_i^{SNU}[t]^{-\alpha} + P_{i, \text{NLoS}}[t] \kappa \beta_0 d_i^{SNU}[t]^{-\alpha}.
\end{equation}

\par The achievable rate between sensor $i$ and the UAV during time slot $t$ is given by
\begin{equation}
R_i[t] = B_i[t] \log_2\left(1 + P_i^{H}[t] |h_i[t]|^2 /\sigma^2 \right),
\end{equation}
where $\sigma^2$ denotes the noise power at the receiver, and $B_i[t]$ represents the communication bandwidth allocated to the BLS $i$ during time slot $t$. Therefore, the amount of data collected by the UAV in time slot $t$ can be expressed as $D[t] = \sum_{i=1}^{N_{SN}} b[i] R_i[t] t_d$, where $b[i]$ is equal to 1 if $d_i^{SNU}[t] \leq D_{\max}$, and 0 otherwise.

\par The UAV dynamically adjusts its position to improve data collection, and we describe its mobility and energy consumption models in the following section.

%
\subsection{UAV Mobile and Energy Cost Models} 
\par In each time slot, the UAV executes a movement action denoted by $\mathbf{a}[t] = [a^x[t], a^y[t], 0]$. Consequently, the position of UAV at time slot $t$ is updated as $\mathbf{c}^{U}[t] = \mathbf{c}^{U}[t-1] + \mathbf{a}[t]$. The energy usage of UAV is influenced by two main components: energy expended for WPT and energy consumed for propulsion. In particular, the charging energy required by the UAV for powering the BLSs is calculated as $E^C[t] = P_\mathrm{w}^U[t] t_d$.
\par Next, we introduce the propulsion energy consumption model for the UAV. For a rotary-wing UAV flying in 3D space, the energy consumption at time slot $t$, denoted as $E^P[t]$, follows the model presented in~\cite{Pan2023}.
\par In summary, the energy consumption of UAV consists of charging and propulsion energy consumption in time slot $t$ can be given by $E[t] = \left(E^C[t] + E^P[t]\right)$.

%
\subsection{Problem Formulation} 
\par This work focuses on two main objectives of maximizing the collected data volume and minimizing the energy consumption of the UAV. In scenarios with multiple users, the UAV may serve only a small subset of BLSs, which can cause uneven data collection. Moreover, some sensors may gather more data than others, resulting in unequal service quality across the system. Therefore, we aim to define a new objective function that jointly optimizes the collected data volume, fairness index, and UAV energy consumption

\par To measure fairness, we adopt Jain’s fairness index~\cite{Masaracchia2020}, which ensures that all sensors achieve reasonable levels of data collection. This index evaluates the balance in energy distribution and reduces disparities in data collection across sensors. The fairness index is given by
\begin{equation}
F[t]=\frac{\left(\sum_{i=1}^{N_{S N}} \sum_{k=1}^t P_i^H[k] \cdot t_d\right)^2}{N_{S N} \sum_{i=1}^{N_{S N}}\left(\sum_{k=1}^t P_i^H[k] \cdot t_d\right)^2},
\end{equation}
where \(F[t] \in [0, 1]\). A higher fairness index indicates more equal energy distribution and ensures a fair data collection process.

\par Based on this, the transmit power and the trajectory of the UAV should be jointly controlled to maximize the collected data volume and fairness index while minimizing the UAV energy consumption. The optimization requires the determination of two decision variables: (i) $\mathbf{P}=\left\{P_{\mathrm{w}}^U[t] \mid t \in \mathcal{T}\right\}$, which is a matrix that defines the transmit power of the UAV at each time slot, and (ii) $\mathbf{A}=\left\{\left[a^x[t], a^y[t], 0\right] \mid t \in \mathcal{T}\right\}$, which is a matrix that defines the spatial displacement of the UAV over time.

\par As such, the joint optimization problem can be expressed as follows:
\begin{subequations} \label{eq:joint_opt}
\begin{align}
\max _{\mathbf{P},\mathbf{A}} 
& \quad f = \sum_{t=1}^{T} \left( -\alpha E[t] + \beta D[t] + \gamma F[t] \right), \\
\text{s.t.} 
& \quad C1: X_{\min} \leq x^U[t] \leq X_{\max}, \quad \forall t, \label{eq:C1} \\
& \quad C2: Y_{\min} \leq y^U[t] \leq Y_{\max}, \quad \forall t, \label{eq:C2} \\
& \quad C3: \left\|a^x[t]\right\| \leq x_{\max}, \quad \forall t, \label{eq:C3} \\
& \quad C4: \left\|a^y[t]\right\| \leq y_{\max}, \quad \forall t, \label{eq:C4} \\
& \quad C5: P_\mathrm{min}^{U} \leq P_\mathrm{w}^{U}[t] \leq P_\mathrm{max}^{U}, \quad \forall t, \label{eq:C5} \\
& \quad C6: R_{i}[t] \geq R_{th}, \quad \forall t. \label{eq:C6}
\end{align}
\end{subequations}
where (\ref{eq:C1}) and (\ref{eq:C2}) indicate that the UAV can only fly within the designated target area. Moreover, the movement distance of the UAV in each time slot is constrained by (\ref{eq:C3}) and (\ref{eq:C4}). In addition, (\ref{eq:C5}) limits the transmit power for charging the UAV. The parameters $\alpha$, $\beta$, and $\gamma$ are weight factors that represent the relative importance of the corresponding terms. The parameter $R_{th}$ represents the minimum achievable rate. Furthermore, (\ref{eq:C6}) specifies that the communication rate must satisfy the minimum rate constraint $R_{th}$. As noted in \cite{He2023}, this problem is non-convex and generally quite difficult to solve in real-time.

%
\section{DRL-based approach} 
\label{sec:DRL_based_approach}
\par 
In this section, we propose a DRL-based approach to solve our joint optimization problem. To this end, we first show the motivations for using DRL and reformulate the problem as a Markov decision process (MDP). Then, we introduce the proposed SAC-PP algorithm with several improvements.

\begin{figure}
  \centering\includegraphics[width=3.5in]{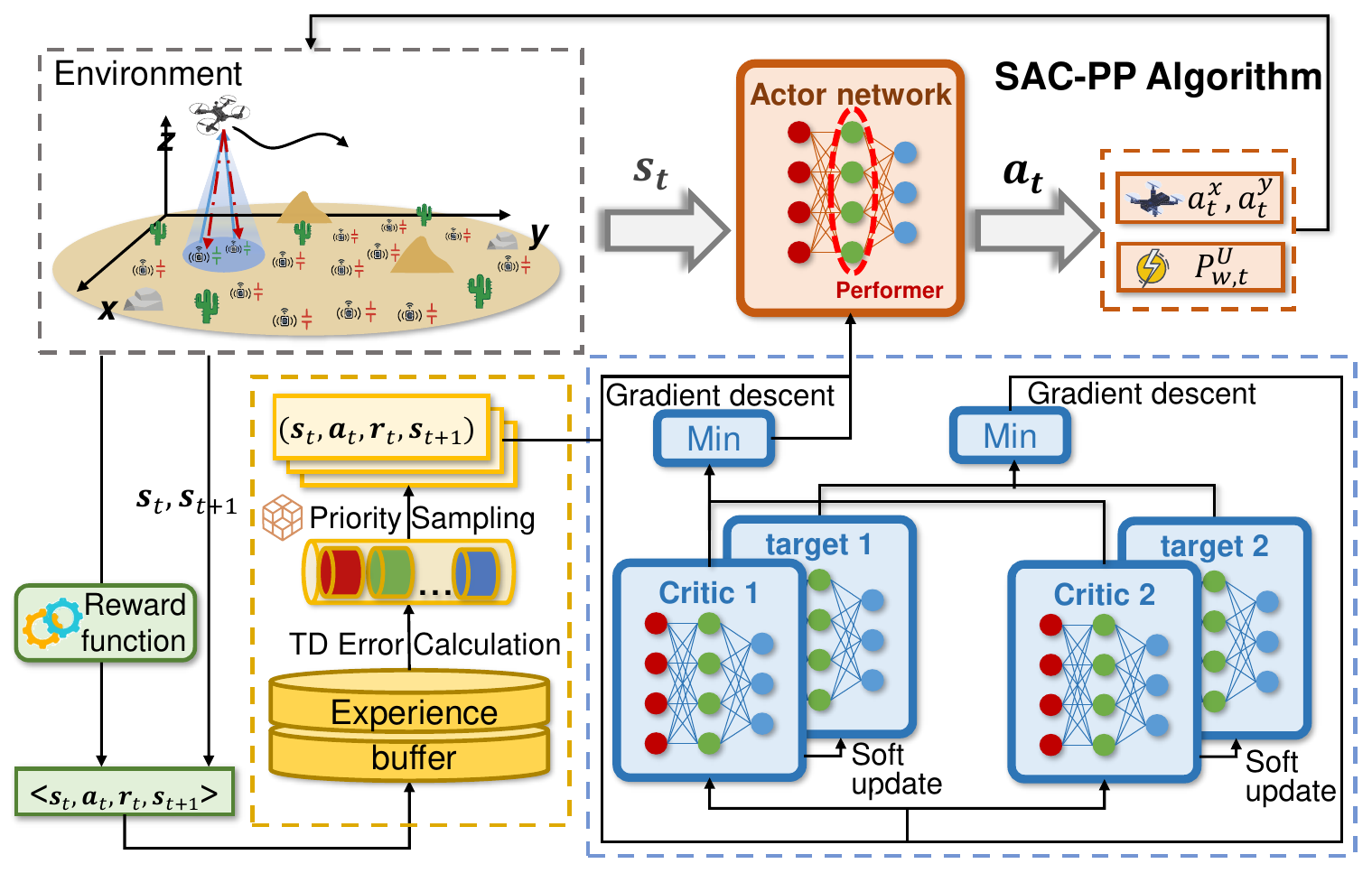}
  \caption{The framework of the SAC-PP algorithm.}
  \label{fig:SAC-PP}
\end{figure}

\subsection{MDP Formulation}
\label{ssec:DRL}
\par 
The considered joint optimization problem is dynamic and uncertain, with unpredictable factors such as sensor data demands and UAV mobility that can change rapidly. Traditional optimization methods struggle to handle these complexities effectively\cite{Li2024}. In contrast, DRL can adapt to changing conditions and offer reliable solutions. Thus, we adopt the DRL approach to solve this challenging optimization problem.

\par 
To address this, we transform the original optimization problem in Eq. \eqref{eq:joint_opt} into an MDP framework. An MDP is characterized by the elements $(\mathcal{S}, \mathcal{A}, \mathcal{P}, R, \gamma)$, where $\mathcal{S}$ denotes the state space, $\mathcal{A}$ represents the action space, $\mathcal{P}$ defines the transition probabilities, $R$ is the reward function, and $\gamma$ stands for the discount factor. Among these, the state, action, and reward are key components, which is detailed as follows.

\begin{enumerate}[label=\textbf{1)}, left=0pt]
    \item \textbf{State Space:} Given that the agent can monitor the UAV and BLS conditions, the state $s[t] \in \mathcal{S}$ is given by
    \begin{align}
    s[t]=\left\{\mathbf{c}^U[t],\mathbf{c}_1^{SN},...,\mathbf{c}_{N_{SN}}^{SN}, \mathbf{f}_1^{SN}[t],...,\mathbf{f}_{N_{SN}}^{SN}[t] \right\},
    \end{align}
    where $\mathbf{f}_i^{SN}[t]$ is the data uploading status of the BLS $i$ in time slot $t$. Specifically, when sensor $i$ has completed all data uploads, $\mathbf{f}_i^{S N}[t]=1$, otherwise, $\mathbf{f}_i^{S N}[t]=0$.
\end{enumerate}
\begin{enumerate}[label=\textbf{2)}, left=0pt]
    \item \textbf{Action Space:} In time slot $t$, the agent selects the action $a[t] \in \mathcal{A}$ based on the state $s[t]$, as follows:
    \begin{align}
    a[t]=\left(a^x[t], a^y[t], P_{\mathrm{w}}^U[t]\right),
    \end{align}
    where $a^x[t], a^y[t]$ are the flight distance of UAV along the $x$, $y$ axes. $P_{\mathrm{w}}^U[t]$ is the transmit power of the UAV in time slot $t$.
\end{enumerate}
\begin{enumerate}[label=\textbf{3)}, left=0pt]
    \item \textbf{Reward Function:} The reward function has a significant impact on the behavior of the agent, so it is designed to align with the optimization objectives. Specifically, the reward $r[t] \in \mathcal{R}$ is given by
    \begin{align}
    r[t] = r^E[t] + r^D[t] + r^F[t] + r^P[t], \label{eq:reward}
    \end{align}
    where $r^E[t] = -\alpha E[t]$, $r^D[t] = \beta D[t]$, and $r^F[t] = \gamma F[t]$ correspond to the three optimization objectives, respectively. Additionally, $r^P[t]$ represents a penalty applied when the UAV exits the target area.
\end{enumerate}

\subsection{SAC Algorithm}
\label{ssec:SAC}
\par  We adopt the Soft Actor Critic (SAC) algorithm~\cite{Wang2024} to solve the MDP. SAC maximizes the expected return and the policy entropy to encourage exploration and improve robustness. The objective is given by
\begin{align}
J(\pi) = \mathbb{E}{\tau \sim \pi} \left[ \sum\nolimits{t=0}^T r_t + \alpha \mathcal{H} \left( \pi \left( \cdot \mid s_t \right) \right) \right], \label{eq:entropy_objective}
\end{align}
where $\mathcal{H} \left( \pi \left( \cdot \mid s_t \right) \right)$ is the policy entropy and $\alpha$ balances exploration and exploitation.
\par  SAC uses an actor-critic architecture with two Q-networks to reduce estimation bias and a target value network to stabilize training. The target network is updated by a soft update mechanism. The optimization objective is given by
\begin{align}
J(\theta, \pi) = \mathbb{E}{\tau \sim \pi} \left[ \sum\nolimits{t=0}^T Q_\pi \left( s_t, a_t \right) - \alpha \log \pi \left( a_t \mid s_t \right) \right], \label{eq:sac_objective}
\end{align}
where $Q_\pi(s_t, a_t)$ is the action value estimate and $\alpha \log \pi(a_t \mid s_t)$ encourages exploration through entropy maximization.
\par  
However, SAC can be inefficient in high dimensional state and action spaces because complex dependencies limit the effective use of experience samples, which slows convergence and may lead to suboptimal policies. Therefore, we enhance SAC to improve performance in complex environments.

\subsection{SAC-PP}
\label{ssec:SAC-PP}
\par This subsection presents an enhanced version of the SAC algorithm, namely, SAC-PP. Specifically, SAC-PP improves algorithm performance and convergence speed by incorporating two key enhancements, which are PER and performer, and they are detailed as follows.

\par \textit{1) PER:} In standard SAC, transitions are sampled uniformly from the replay buffer, which can under sample informative experiences and reduce training efficiency. Moreover, the finite buffer capacity may overwrite valuable transitions before they are sufficiently reused, which can degrade stability. Therefore, we adopt prioritized experience replay (PER)~\cite{Li2025a}, which samples transitions according to their temporal difference (TD) error. The TD error is given by
\begin{equation}
\delta_i = \left| r_t + \gamma Q(s_{t+1}, a') - Q(s_t, a_t) \right|.
\end{equation}
The sampling probability is defined as $P(i) = p_i^\alpha / \sum\nolimits_j p_j^\alpha$, where $p_i$ is typically set proportional to $\delta_i$ and $\alpha$ controls the prioritization strength. To reduce the bias introduced by non uniform sampling, importance sampling weights are applied in gradient updates. The weight is given by $w_i = \left( 1 /(N P(i)) \right)^\beta$, where $N$ is the buffer size and $\beta$ controls the degree of correction.

\par \textit{2) Performer:} Learning long horizon strategies is challenging because the temporal coupling between power control and trajectory planning requires modeling dependencies across many time steps. Attention based architectures can capture such dependencies, but standard self attention is expensive for long sequences. We adopt performer~\cite{Choromanski2021}, which approximates self attention using random feature based kernel mapping and achieves linear complexity with respect to the sequence length. The attention computation is given by
\begin{equation}
\operatorname{Attention}(Q, K, V) = \phi(Q) \left(\phi(K)^T V\right),
\end{equation}
where $Q$, $K$, and $V$ are the query, key, and value matrices, and $\phi(\cdot)$ maps queries and keys into a feature space that enables efficient computation. We integrate the performer module into the actor network to strengthen sequential representation learning under limited computational budget.

\begin{algorithm}[!b]
  \caption{SAC-PP}\label{algo:SAC-PP}
    Initialize the parameters $\theta$ of the actor network, $\phi$ of the critic network, and $\alpha$ of the entropy coefficient;\\
    Initialize the target critic network parameters $\phi_{\text{target}} \leftarrow \phi$;\\
    Initialize the PER buffer $C$;\\

  \For{Episode $= 1, \ldots, N^{eps}$}
  {
    Reset the environment and initialize state $s_t$;\\
    \For{Time step $t = 1, \ldots, T$}
    {
        Obtain state $s[t]$;\\
        Select action $a[t]=\{a^x[t], a^y[t], P_{\mathrm{w}}^U[t] \}$;\\
        Execute action $a[t]$;\\
        Calculate the collected data volume of the UAV $D[t]$, the UAV energy consumption $E[t]$, and the fairness index $F[t]$;\\
        Calculate r[t] according to Eq. ~\eqref{eq:reward};\\
        Store the transition $(s[t], a[t], r[t], s[t+1])$ in PER buffer $C$ with priority $p_i$;\\

        \If{Size(buffer) $\geq$ batch size}
        {  
            Retrieve a prioritized mini-batch of transitions based on $p_i$;\\  
            Update the critic by minimizing the loss as described in Eq.~\eqref{eq:sac_objective};\\  
            Optimize the actor by maximizing the objective from Eq.~\eqref{eq:entropy_objective};\\
            Adjust the entropy coefficient $\alpha$ via gradient descent;\\  
            Soft update the target critic network:  
            \[
            \phi_{\text{target}} \leftarrow \tau \phi + (1 - \tau) \phi_{\text{target}};
            \]
        }
    }
  }
\end{algorithm}

\par \textit{3) Main Steps of SAC-PP Algorithm:} As shown in Fig.~\ref{fig:SAC-PP}, SAC-PP applies an actor network to learn the policy and a critic network for value estimation. In state $s_t$, the actor network, enhanced with the performer module, outputs action $a_t$. Upon execution, the environment provides a reward $r_t$, creating a state transition. This transition, along with the TD error, is processed by the PER module and added to the experience replay buffer for further learning.

\par Algorithm~\ref{algo:SAC-PP} outlines SAC-PP. Initially, the actor and critic networks and the PER buffer are initialized. For every episode, environment and state reset. At each time step, an action is chosen and executed, and the collected data volume, fairness index, and UAV energy consumption are computed, followed by receiving reward $r_t$. The transition and TD error then go into the buffer. Once the buffer reaches the batch size, the networks undergo multiple training epochs to ensure convergence. This cycle continues across episodes, progressively refining the policy.

%
\section{SImulation Results and Analysis} 
\label{sec:SImulation_Results_and_Analysis}
\par In this section, we first introduce the simulation setting and benchmarks, and then provide the simulation results.

\subsection{Simulation Setups}
\label{ssec:Simulation_Setups}


\begin{figure*}[htbp]
    \centering
    \subfloat[]{
        \includegraphics[width=0.336\linewidth, keepaspectratio]{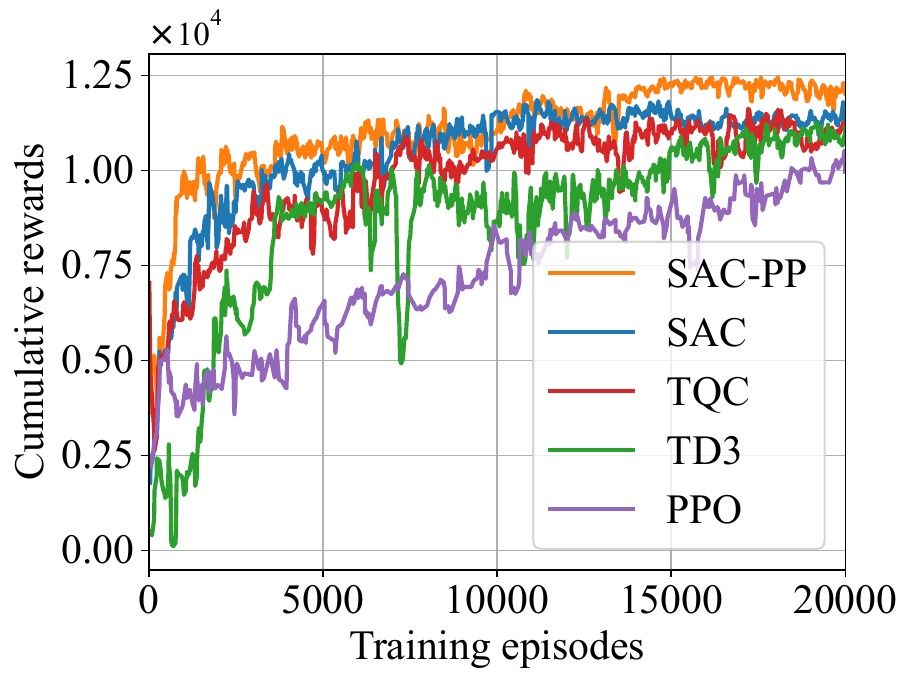}\label{fig:reward}}
    \hspace{0\linewidth}
    \subfloat[]{
        \includegraphics[width=0.18\linewidth, keepaspectratio]{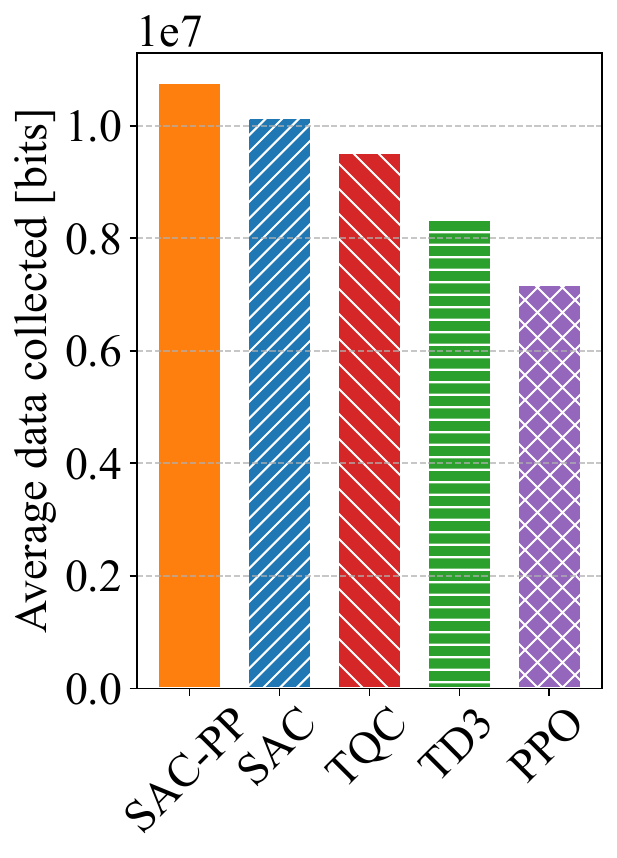}\label{fig:data}}
    \hspace{0.01\linewidth}
    \subfloat[]{
        \includegraphics[width=0.18\linewidth, keepaspectratio]{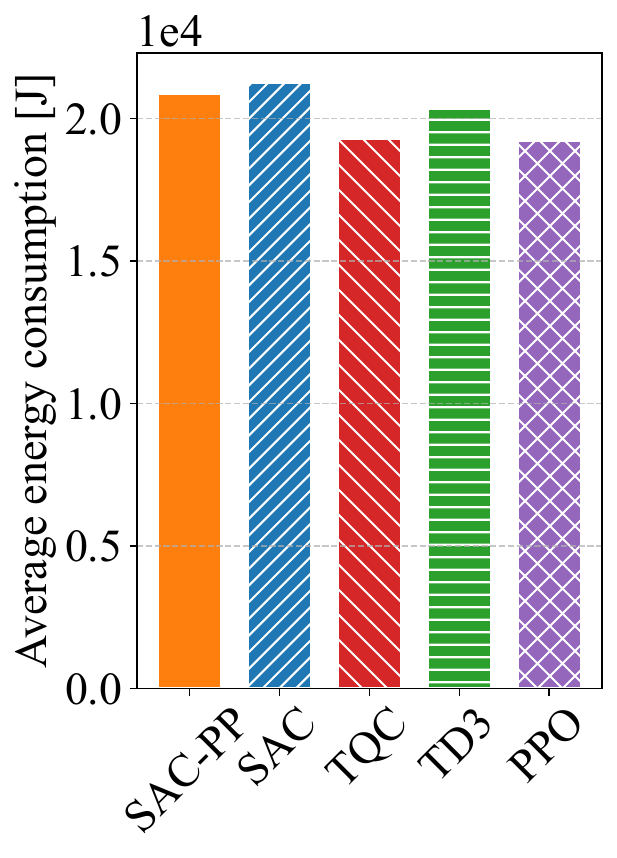}\label{fig:energy}}
    \hspace{0.01\linewidth}
    \subfloat[]{
        \includegraphics[width=0.18\linewidth, keepaspectratio]{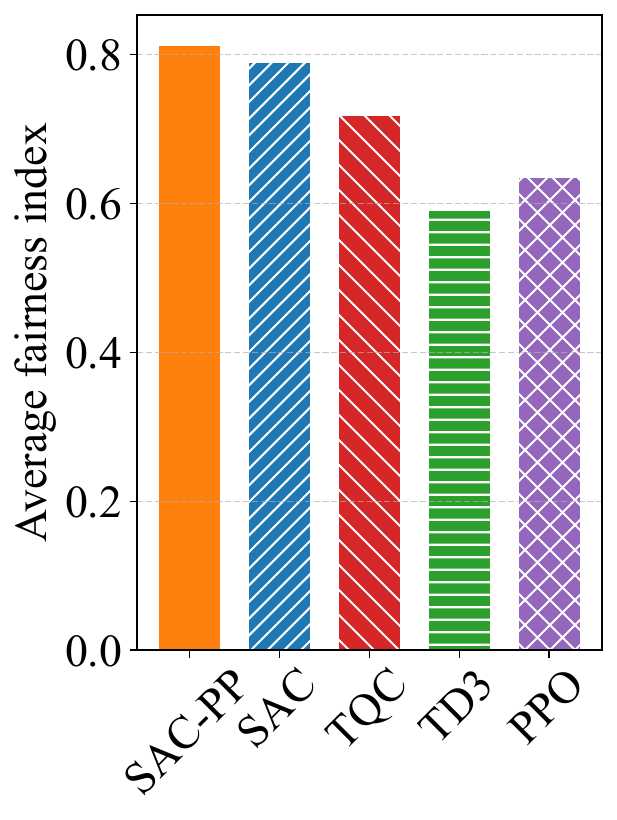}\label{fig:fairness}}
    \caption{(a) Cumulative rewards training curve. (b) Average data collected [bits]. (c) Average energy consumption [J]. (d) Average fairness index.}
    \label{fig:results}
\end{figure*}

\par 
We consider a target area of size $400 \mathrm{m} \times 400 \mathrm{m}$, where a UAV serves 20 BLSs randomly distributed within the region. The UAV operates within this area at a fixed altitude of 50 m. We set the bottom-left corner of the monitoring area as the origin, and the initial coordinates of the UAV are $(100,100)$. In each time slot, the data generation state of the sensors is refreshed, with the occurrence of new data governed by a Bernoulli distribution. When new data is generated, the volume of the data is modeled by a normal distribution. Other parameters related to communication follow references~\cite{Panahi2024}~\cite{Pan2023}~\cite{Wang2023}.
\par 
For comparison, we adopt the twin delayed deep deterministic policy gradient (TD3)~\cite{Wang2024}, twin critic with quantile regression (TQC)~\cite{Kuznetsov2020}, proximal policy optimization (PPO)~\cite{Wang2024}, and soft actor-critic (SAC) as benchmark methods~\cite{Wang2024}.

\subsection{Optimization Results}
\label{ssec:Optimization_Results}

\par Fig.~\ref{fig:results}(a) illustrates cumulative rewards for SAC-PP against other algorithms. Initially, all algorithms show unstable rewards due to poor performance. As they converge, rewards stabilize, indicating strategy identification. SAC-PP shows reward improvements during training due to energy consumption. Furthermore, SAC-PP outperforms others in reward and stability for two reasons. First, the Performer module improves temporal modeling. Second, the PER module prioritizes samples with high TD errors, improving training efficiency. Thus, SAC-PP maximizes collected data volume and fairness index while minimizing energy consumption.

\par Fig.~\ref{fig:results}(b) compares average collected data volume across five methods. SAC-PP achieves the highest collected data volume, surpassing other methods, while TD3 and PPO perform weakest. SAC-PP explores complex state spaces and utilizes data more efficiently. PER prioritizes samples with high TD errors, improving training efficiency. Meanwhile, Performer applies linearized self-attention to capture temporal dependencies, enhancing data collection in dynamic environments. Together they enable efficient learning and policy optimization over time horizons.

\par Fig.~\ref{fig:results}(c) compares average energy consumption of five methods. SAC-PP maintains a high level of collected data while keeping energy consumption comparable to SAC and TQC, indicating a good balance. PPO achieves lowest energy but sacrifices collected data volume. TD3 and TQC show moderate energy consumption but deliver weaker performance. SAC-PP reduces energy consumption through precise policy updates. PER prioritizes experiences for efficient learning, while Performer captures dependencies between states.

\par Fig.~\ref{fig:results}(d) shows fairness index comparison among five methods. SAC-PP achieves highest fairness, reflecting improved resource coordination in multi-user scenarios. SAC and TQC maintain relatively high fairness. In contrast, TD3 and PPO show lower fairness indices, indicating imbalanced resource allocation that affects system stability. SAC-PP considers user and task needs in decisions. PER helps reveal issues in resource allocation, while Performer captures long-term dependencies across time sequences.

%
\section{Conclusion} 
\label{sec:Conclusion}
This paper explored a UAV-assisted joint data collection and WPT system for BLS networks. Specifically, the transmit power and trajectory of the UAV are optimized to maximize the collected data volume and fairness index while minimizing the UAV energy consumption. Then, we propose a SAC-PP algorithm, which integrates PER and the performer module to enhance stability and convergence speed. Simulation results indicate that SAC-PP performs better than other baseline methods, particularly in optimizing the collected data volume, fairness index, and UAV energy consumption.

\section*{Acknowledgment}
This research is supported in part by the National Natural Science Foundation of China (62272194, 62471200), and in part by the Science and Technology Development Plan Project of Jilin Province (20250101027JJ), in part by Seatrium New Energy Laboratory, Singapore Ministry of Education (MOE) Tier 1 (RT5/23 and RG24/24), the Nanyang Technological University (NTU) Centre for Computational Technologies in Finance (NTU-CCTF), and the Research Innovation and Enterprise (RIE) 2025 Industry Alignment Fund - Industry Collaboration Projects (IAF-ICP) (Award I2301E0026), administered by Agency for Science, Technology and Research (A*STAR), in part by the Postdoctoral Fellowship Program of China Postdoctoral Science Foundation (GZC20240592), in part by China Postdoctoral Science Foundation General Fund (2024M761123), and in part by the Scientific Research Project of Jilin Provincial Department of Education (JJKH20250117KJ).


\bibliographystyle{IEEEtran}
\bibliography{ref}

\end{document}